# Superconductivity in defective pyrite-type iridium chalcogenides Ir$_x$Ch$_2$ (Ch = Se and Te)

Yanpeng Qi,[1] Satoru Matsuishi,[2] Jiangang Guo,[1] Hiroshi Mizoguchi,[1] and Hideo Hosono[1,2,*]

1. Frontier Research Center, Tokyo Institute of Technology, 4259 Nagatsuta, Midori, Yokohama 226-8503, Japan
2. Materials and Structures Laboratory, Tokyo Institute of Technology, 4259 Nagatsuta, Midori, Yokohama 226-8503, Japan

**Abstract**

We report superconductivity in defective pyrite-type iridium chalcogenides Ir$_x$Ch$_2$ (Ch = Se and Te). Maximum values of $T_c$ of 6.4 K for Ir$_{0.91}$Se$_2$ and 4.7 K for Ir$_{0.93}$Te$_2$ were observed. It was found that Ir$_{0.75}$Ch$_2$ (Ir$_3$Ch$_8$) is close to the boundary between metallic and insulating states and Ir$_x$Ch$_2$ systems undergo nonmetal to metal transitions as $x$ increases. On the basis of density functional theory calculations and the observed large variation in the $Ch-Ch$ distance with $x$, we suggest that Ir$_{0.75}$Ch$_2$ (Ir$_3$Ch$_8$) is the parent compound for the present superconductors.

[*] Author to whom correspondence should be addressed. E-mail: hosono@msl.titech.ac.jp

Superconductivity often emerges when an ordered state disappears as a result of a perturbation such as carrier doping or pressure. High-$T_c$ cuprates appear between a Mott insulating phase with antiferromagnetic ordering and a conventional metallic phase with doping carriers [1]. Superconductivity in iron pnictides appears between an antiferromagnetic metallic phase with spin-density wave ordering and a Fermi metallic phase [2]. These results indicate that it is essential to choose an appropriate parent material for exploring novel superconductors.

In this Letter, we report that the nearly insulating iridium chalcogenides $Ir_{0.75}Ch_2$ ($Ir_3Ch_8$; $Ch$ = Se, Te) [3,4] could be the parent compounds of superconductors, along with our experimental findings. These compounds have a pyrite structure and belong to the space group $Pa\bar{3}$ (see **Fig. 1(a)**). Each unit cell contains four Ir and eight $Ch$ sites, and Ir sites with randomly distributed vacancies form a simple face-centered-cubic (fcc) lattice. The $Ch$ atoms form four $Ch_2$ chalcogen dimers centered at the midpoints of the cube edges and at the body center, so the structure may be considered to be rock-saltlike. The $Ch-Ch$ bonds of dimers are aligned along the body diagonals of the cubic cell. Each Ir is octahedrally coordinated by six $Ch$ atoms, and each $Ch$ is tetrahedrally coordinated by three Ir atoms and one $Ch$ atom; a three-dimensional bond sequence $-Ch-Ch-Ir-Ch-Ch-Ir-Ch-Ch-$ is formed. In contrast to stoichiometric pyrite-type platinum-group chalcogenides, which generally have metallic conductivity, $Ir_3Te_8$ has relatively high electrical resistivity (~0.7 mΩ cm at room temperature), with a negative temperature coefficient of resistivity in low-temperature regions [5], implying that this compound is close to the boundary between metallic and insulating states. In the course of our search for new superconductors, we performed the high-pressure synthesis of pyrite-type $Ir_xCh_2$ with $x$ close to 1, and found superconductivity with maximum $T_c$ values of 6.4 K and 4.7 K for $Ir_{0.91}Se_2$ and $Ir_{0.93}Te_2$, respectively. Density functional theory (DFT) calculations performed on $Ir_3Se_8$ with ordered Ir vacancies and $IrSe_2$ structures revealed that the Fermi level in pyrite-type $Ir_xSe_2$ is composed of antibonding σ orbitals (σ*) of the Se−Se dimer and $dz^2$ orbitals of the Ir. All Se dimers have equal bond lengths in $IrSe_2$, so they form wide conduction bands. It is therefore considered that the separation and narrowing of the conduction bands in $Ir_3Se_8$ are caused by the coexistence of long and short Se−Se dimers. This result implies that the nonmetal to metal transition and subsequent metal to superconductor transition are driven by relaxation as a result of alternating bond-lengths in the array of Se dimers.

Under ambient conditions, $IrCh_2$ compounds have a crystal structure different from that of pyrite [6,7]. A high-pressure synthetic technique was therefore used, on the basis of a literature report of the stabilization of $Ir_xTe_2$ pyrite under high pressure [8]. Fine powders of Ir (99.9wt.%) and Se (99.9wt.%) or Te (99.9wt.%), were mixed in a desired ratio and were placed in an $h$-BN capsule, and then heated at 1673 K and 5 GPa for 2 h using a belt-type high-pressure apparatus. All starting materials and precursors for the synthesis were prepared in a glove box filled with purified Ar gas ($H_2O$, $O_2$ < 1 ppm).

The crystalline phases in the resulting samples were identified by powder X-ray diffraction (XRD) using a Bruker diffractometer model D8 ADVANCE (Cu rotating anode). Rietveld refinement of the XRD patterns was performed using the code TOPAS3 [9]. Elemental compositions were determined using an electron-probe micro-analyzer (EPMA; JEOL, Inc., model JXA-8530F). The micrometer-scale compositions within the main phase were probed at five to ten focal points, and the results were averaged. The dependence of the dc electrical resistivity ($\rho$) on temperature was measured over 2–300 K using a conventional four-probe method. Magnetization ($M$) measurements were performed using a vibrating sample magnetometer (Quantum Design). Specific-heat data were obtained using a conventional thermal relaxation method, using Quantum Design PPMS.

The DFT calculations were performed using the generalized gradient approximation (GGA) with the Perdew–Burke–Ernzerhof (PBE) functional and the projected augmented plane-wave method implemented in the Vienna *ab initio* simulation program (VASP) [10–12]. Spin–orbit coupling of the valence electrons was included using the second-variation method using the scalar-relativistic eigenfunctions of the valence states [13]. The lattice parameters and atomic positions were fully relaxed by a structural optimization procedure, minimizing the total energy and force. Self-consistent solutions of the Kohn–Sham equations were obtained using a $4 \times 4 \times 4$ Monkhorst–Pack grid of $k$-points for the integration over the Brillouin zone and the plane-wave basis-set cutoff was set to 600 eV. The full electronic density of states (DOS) was also calculated using a high $k$-point sampling of $8 \times 8 \times 8$. The projected DOS (PDOS) of each atom was obtained by decomposing the charge density over the atom-centered spherical harmonics with a Wigner–Seitz radius of 1.423 Å for Ir and 1.164 Å for Se.

Chemical composition analysis using EPMA demonstrated that the one-to-one correspondence between actual and nominal compositions at $x < 0.9$ is good. With increasing nominal $x$, the segregation of Ir metal and the saturation of analyzed $x$ occurred at around $x \geq 0.91$ for $Ir_xSe_2$ and $x \geq 0.93$ for $Ir_xTe_2$. **Fig. 1(b)** shows the XRD pattern of the $Ir_{0.91}Se_2$ sample, representing pyrite-type $Ir_xSe_2$. Except for the small peaks arising from Ir metal impurity, all peaks could be attributed to a pyrite-type structure (cubic, space group $Pa3$, $Z = 4$) [14–16]. With increasing Ir content, the geometrical distortion of $IrSe_6$ octahedra rapidly increased (see Supplementary

Materials). The lattice parameter $a$ increased linearly with $x$ from 5.947 Å at $x = 0.68$ to 5.980 Å at $x = 0.91$. At the same time, the bond length of the Se dimer ($r_{Se–Se}$) significantly increased from 2.423 Å to 2.652 Å (see **Fig. 1c**). The rate of increase (9.43%) of $r_{Se–Se}$ is an order of magnitude higher than that (0.56%) of $a$, indicating that the bonding state in the Se dimers is significantly controlled by the Ir content. A similar phenomenon was also observed in the $Ir_xTe_2$ system.

**Figure 2** shows the temperature dependence of the electrical resistivity for the $Ir_xSe_2$ compounds in the temperature range 2–300 K. The $\rho(T)$ data of $Ir_{0.68}Se_2$ indicate semiconducting behavior with a room-temperature resistivity of 14 Ω cm. With increasing $x$, the room temperature resistivity gradually decreased and a sharp drop in $\rho(T)$ to zero was observed at 3.2 K for $x = 0.77$, suggesting the onset of a superconducting transition. The temperature for zero resistivity increased with increasing $x$ to $T_c = 6.4$ K for $Ir_{0.91}Se_2$. Large diamagnetic signals were clearly observed below the zero resistivity temperature, as shown in **Fig. 2(b)** and **(c)**, indicating the bulk superconductivity of $Ir_xSe_2$. The magnetization curve in the superconducting state showed the typical behavior of type-II superconductors. The value of $\mu_0H_{c2}(0)$ was estimated to be 14.3 T [17], which yields a Ginzburg–Landau coherence length $\xi_{GL}(0)$ of ~48 Å. Further evidence for bulk superconductivity was obtained from the large specific-heat jump at $T_c$, shown in **Fig. 2(d)**. We therefore fitted $C_p(T)/T$ vs $T^2$ with $C_p(T)/T = \gamma + \beta T^2$, which yielded the electronic specific-heat coefficients $\gamma = 6.8$ mJ mol$^{-1}$ K$^{-2}$ and $\beta = 0.376$ mJ mol$^{-1}$ K$^{-4}$. The Debye temperature $\Theta_D = (12\pi^4NR/5\beta)^{1/3}$ was ~249 K. Using $\gamma = 6.8$ mJ mol$^{-1}$ K$^{-2}$, the normalized specific-heat jump value $\Delta C/\gamma T_c$ was estimated to be 3.1, which was substantially larger than the Bardeen–Cooper–Schrieffer (BCS) weak coupling limit value of 1.43 [18], indicating strong coupling superconductivity. The superconducting transitions of $Ir_{0.93}Te_2$ are also shown in **Fig. 2(b)–(d)**, and the relevant parameters are summarized in **Table I**.

**Figure 3** shows the electronic phase diagram of $Ir_xSe_2$. For $x \leq 0.77$; one observes a dramatic decrease in $\rho$ at 7 K, i.e., the resistivity in the normal state just above the superconducting transition, by more than six orders of magnitude. It is clear that the $Ir_xSe_2$ compounds undergo a nonmetal to metal transition. Zero resistivity starts to appear at $x \geq 0.79$ and the onset $T_c$ monotonically increases with increasing $x$ and reaches a maximum $T_c$ of 6.4 K at $x = 0.91$. Samples with $x \geq 0.91$ could not be synthesized under the present conditions. Almost the same phase diagram was obtained in the $Ir_xTe_2$ system and the maximum $T_c$ was observed in $Ir_{0.93}Te_2$ at ~4.7 K (see Supplementary Materials). The present results strongly suggest that the nearly insulating iridium chalcogenides $Ir_3Ch_8$ ($Ch$ = Se or Te) are the parent compounds of the present superconductors. To investigate the origin of the nonmetal to metal transition coincident with the emergence of superconductivity, we performed DFT calculations. **Fig. 4(a)** shows the structure and electronic structure of a vacancy-ordered structural model of $Ir_3Se_8$ in which one of the four Ir sites in a unit cell of pyrite $IrSe_2$ was removed. As a result of structural optimization minimizing the total energy and force, a rhombohedral structure (space group $R3$, $a$ = 6.066 Å, $\alpha$ = 90.43°, $Z$ = 1) was obtained, with one group of Ir and two groups of Se sites (Se1 and Se2) containing one long Se1−Se1 ($r_{Se1-Se1}$ = 2.61 Å) and three short Se2−Se2 dimers ($r_{Se2-Se2}$ = 2.50 Å) per unit cell (see the left panel in **Fig. 4a**). As shown in the center panel of **Fig. 4(a)**, only one narrow band crosses the Fermi level ($E_F$). The total DOS and the momentum-projected DOS on spherical harmonics centered at Ir ($d$ component), and Se1 and Se2 ($p$ components) sites, shown in the right panel, indicate that the conduction band crossing $E_F$ is mainly composed of Ir 5$d$ and Se1 4$p$ orbitals. The electron density isosurface of the narrow conduction band within the energy region below $E_F$ (gray region in the center panel) is plotted on the structural model in the left panel. It is evident that the σ* orbital in the Se1−Se1 dimer located at the center of the unit cell and the $dz^2$ orbitals of the nearest Ir constitute this band. Here, we note that the contribution from the σ* orbitals of the Se2−Se2 dimers located at the edges of the unit cell is negligibly small, indicating a distinct influence of the Ir vacancy on the σ* orbital level of these dimers. Basically, the $Ch_2$ dimer in a pyrite-type crystal forms a divalent anion $Ch_2^{2-}$, and the electronic configuration of the molecular orbitals is σ(2)π$_x$(2)π$_y$(2)π$_x$*(2)π$_y$*(2)σ*(0). Considering the charge transfer from Ir to the Se1−Se1 dimer, Ir$_3$Se$_8$ may be expressed as Ir$^{(3-\delta/3)+}_3$Se$_2^{(3-\delta)-}$(Se$_2^{2-}$)$_3$. We consider that the elongation of the Se1−Se1 distance lowers the energy of the σ* orbital in the Se1−Se1 dimer, resulting in the formation of a half-filled narrow band (~1 eV) separated from the higher energy bands composed of the σ* orbital on the Se2−Se2 dimers. Such a half-filled metal band is electronically unstable, and is converted into an insulator by additional interactions such as electron–electron correlations (Mott–Hubbard transitions) and electron–lattice interactions (Peierls transitions). In addition, the actual compound possesses the disorder of Ir vacancies, which would make the metallic state unstable.

We also performed DFT calculations on the pyrite-type IrSe$_2$ structure without an Ir vacancy (see **Fig. 4b**). In this case, a cubic structure (space group $P$a3, $a$ = 6.109 Å, $Z$ = 4) composed of equivalent Se−Se dimers with $r_{Se-Se}$ = 2.72 Å was obtained. Each of the σ* orbitals in the Se−Se dimers contributes equally to the formation of wide bands near $E_F$, leading to metallic conduction. This material may therefore be represented as Ir$^{(3-\delta)+}_4$(Se$_2^{(3-\delta)-}$)$_4$. Similar results were obtained for pyrite-type Ir$_3$Te$_8$ and IrTe$_2$. The value of $r_{Se-Se}$ in stoichiometric $IrSe_2$ is 7.5% larger than the mean $r_{Se-Se}$ value (2.53 Å) in ideal $Ir_3Se_8$, but the value of the lattice parameter $a$ is larger by only 0.7% of that in $Ir_3Se_8$. This calculated lattice constant is consistent with the results obtained by XRD analysis on $Ir_{0.68}Se_2$ and $Ir_{0.92}Se_2$. The results of the DFT calculations imply that the nonmetallic state of $Ir_xCh_2$ with $x < 0.75$ is caused by a bond-length alternation in the array of $Ch$ dimers. With increasing $x$, a mixed state of long and short dimers may converge to form a system with dimers of equal bond length. Furthermore, even in the metallic state, it is supposed that there is still modulation of the $Ch$ dimer length and this induces charge fluctuations among $Se_2$ σ* and Ir $dz^2$ orbitals, playing an important role in the emergence of superconductivity.

Recently, the superconductivity of $CdI_2$-type $IrTe_2$ with a maximum $T_c = 3.1$ K was reported [19,20]. Ambient-pressure phase $IrTe_2$ is composed of tetravalent $Ir^{4+}$ and $Te^{2-}$ anions, and partially filled Ir 5d $t_{2g}$* orbitals primarily contribute to $E_F$. In contrast, $E_F$ of pyrite-type $Ir_xCh_2$ is composed of Ir 5d $e_g$* orbitals strongly hybridized with σ* orbitals of the $Ch$ dimer. Charge transfer between $Ir^{2+}$ and $Ch_2^{2-}$ occurs by shifting the σ* level of the $Ch_2^{2-}$ by varying the $Ch$–$Ch$ separation with the Ir vacancy. We consider that the bonding state of the $Ch$ dimers plays an important role in the emergence of superconductivity.

In summary, we found new superconductors $Ir_xSe_2$ and $Ir_xTe_2$ with pyrite structures, containing chalcogen dimer anions. The transitions from metal to superconductor and insulator to metal were observed simultaneously when the Ir content was increased to $x \geq 0.74$. The DFT calculations showed that a half-filled single band crosses the Fermi level in $Ir_3Ch_8$ with a pyrite structure, whereas stoichiometric $IrCh_2$ has wide bands near the Fermi level. On the basis of these results, we suggest that $Ir_3Ch_8$ is the parent compound for these superconductors.

**Acknowledgment**

This work was supported by the Funding Program for World-Leading Innovative R&D on Science and Technology (FIRST), Japan.

TABLE I. Parameters obtained for superconductivity of $Ir_xCh_2$ ($Ch$ = Se, Te).

| Compound | $Ir_{0.91}Se_2$ | $Ir_{0.93}Te_2$ |
| --- | --- | --- |
| $T_c$ / K | 6.4 | 4.7 |
| $H_{c2}$ / T | 14.3 | 4.58 |
| $\xi_{GL}(0)$ / Å | 48 | 69 |
| $\Delta C/\gamma T_c$ | 3.1 | 2.0 |
| $\Theta_D$ / K | 249 | 262 |

**Figure Captions**

**FIG. 1. Crystal structure and powder XRD pattern of Ir$_x$Se$_2$.** (a) Crystal structure of pyrite Ir$_x$Se$_2$. (b) observed and fitted XRD pattern of Ir$_{0.91}$Se$_2$. Rietveld refinement results showed good convergence ($R_{wp}$ = 5.76%, $R_p$ = 4.27%). (c) Se–Se distance in the dimer anions with Ir content.

**FIG. 2. Evolution of superconductivity as a function of Ir content for Ir$_x$Ch$_2$.** (a) Temperature dependence of resistivity $\rho(T)$ of Ir$_x$Se$_2$ compounds. Inset: Enlarged view of low-temperature region, showing superconducting transition. (b) Low-temperature $\rho(T)$ for optimal Ir$_{0.91}$Se$_2$ and Ir$_{0.93}$Te$_2$, which show $T_c$ values of 6.4 K and 4.7 K. respectively. The transition width in each compound is ~0.2 K. (c) dc magnetization data at low temperatures under applied magnetic field of 10 Oe. Large diamagnetic signals were clearly observed below $T_c$, indicating bulk superconductivity in the samples. (d) Temperature dependence of specific heat for Ir$_{0.91}$Se$_2$ (black) and Ir$_{0.93}$Te$_2$ (red).

**FIG. 3. Electronic phase diagram of Ir$_x$Se$_2$.**

**FIG. 4. Electronic structures obtained by DFT calculations.** (a) Ir$_3$Se$_8$ (space group: $R3$, $a$ = 6.066 Å, $\alpha$ = 90.43°, Z = 1) and (b) IrSe$_2$ ($Pa3$, $a$ = 6.109 Å, Z = 4). Isosurface of conduction electron density (2.5 × 10$^{-3}$ $e^-$/Å$^3$) is overlapped on structural model in left panels. Center panels show the band structures plotted along the lines connecting $k$-points: $R_{111}$(1, 1, 1), $\Gamma$(0, 0, 0), $X_{010}$(0, 1, 0), and $M_{110}$(1, 1, 0). Total (black) and momentum-projected DOS on Ir-$d$ (red), Se1-$p$ (blue), and Se2-$p$ (green) spherical harmonics are plotted in right panels. In IrSe$_2$, Se1 and Se2 are equivalent.

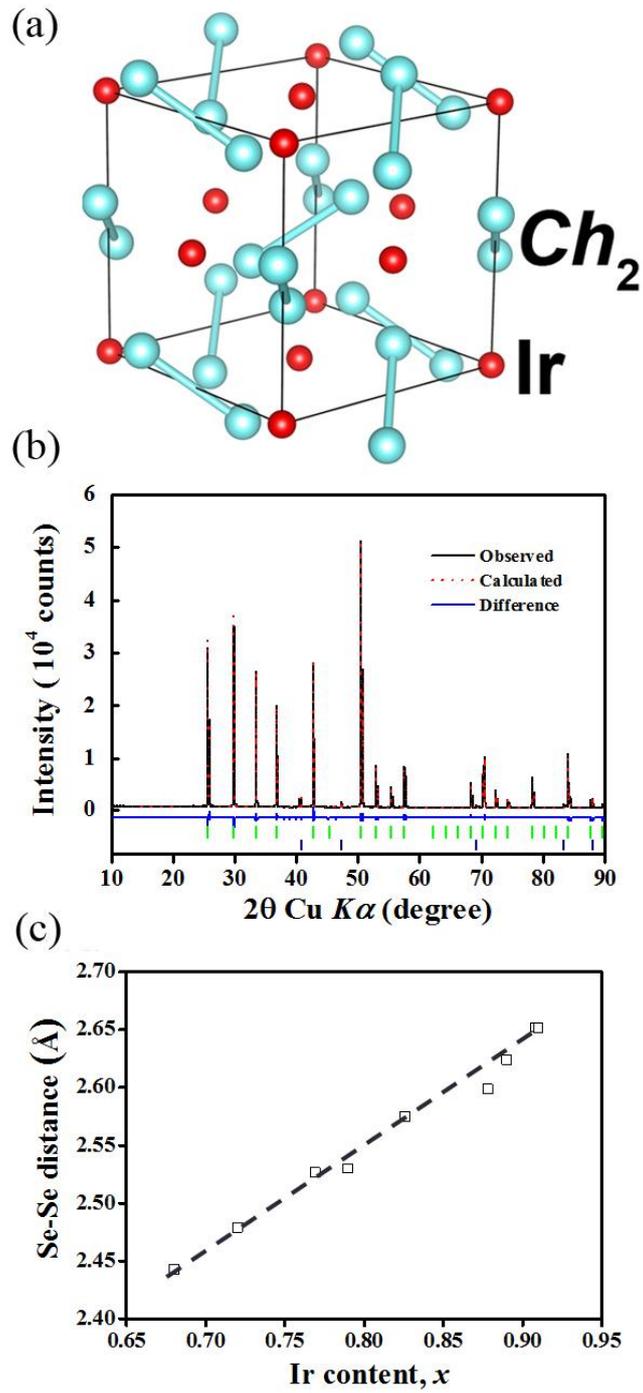

Fig.1 Qi et al.

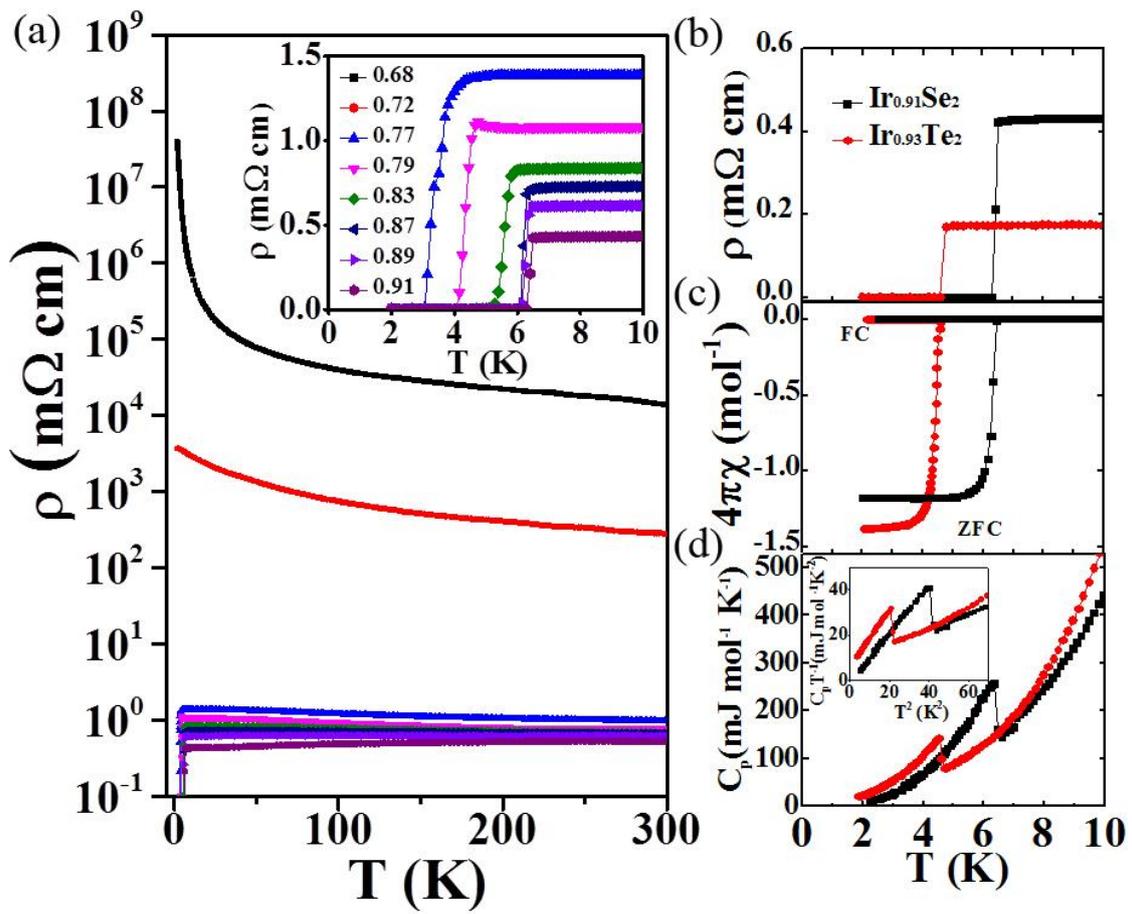

Fig.2 Qi et al.

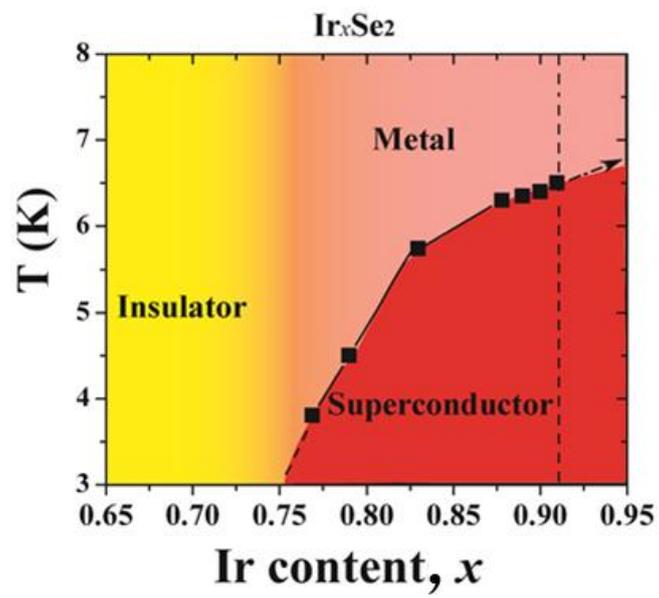

Fig.3 Qi et al.

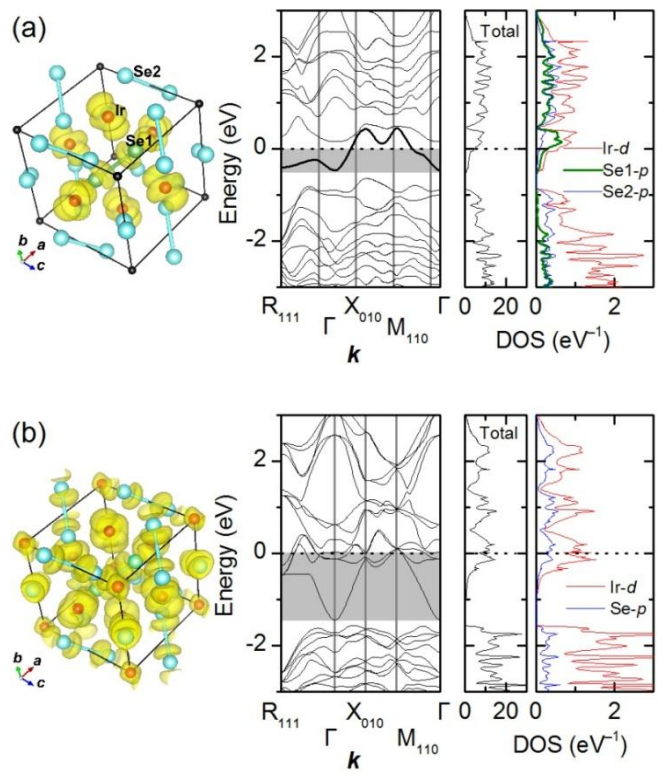

Fig.4 Qi et al.